\documentclass[sigconf]{acmart}
\AtBeginDocument{%
  }

\usepackage{array}

\begin{document}

\title{CapTalk: Unified Voice Design for Single-Utterance and Dialogue Speech Generation}


\author{Xiaosu Su}
\affiliation{%
  \institution{University of Chinese Academy of Sciences}
  \city{Beijing}
  \country{China}}
\email{suxiaosu@iie.ac.cn}

\author{Zihan Sun}
\affiliation{%
  \institution{Hello Group Inc.}
  \city{Beijing}
  \country{China}
}
\email{sun.zihan@hellogroup.com}

\author{Peilei Jia}
\affiliation{%
 \institution{Hello Group Inc.}
 \city{Beijing}
 \country{China}}
\email{jia.peilei@hellogroup.com}

\author{Jun Gao}
\affiliation{%
  \institution{Hello Group Inc.}
  \city{Beijing}
  \country{China}}
\email{gao.jun@hellogroup.com}

\renewcommand\footnotetextcopyrightpermission[1]{} 
\settopmatter{printacmref=false} 

\begin{abstract}
Voice design from natural language descriptions is emerging as a new task in text-to-speech multimodal generation, aiming to synthesize speech with target timbre and speaking style without relying on reference audio. However, existing methods mainly focus on single-utterance generation, leaving conversational voice design largely unexplored. In this work, we extend voice design to dialogue, enabling better target speaker modeling and turn-level expressive control in natural conversational settings. We propose CapTalk, a unified caption-conditioned text-audio autoregressive framework for both single-utterance and dialogue voice design. CapTalk uses utterance-level captions for single-utterance voice design and speaker-level captions for dialogue speaker modeling, and further introduces a CoT control sequence in dialogue to explicitly plan turn-level dynamic attributes. To resolve the conflict between stable timbre preservation and context-adaptive expression, we propose a hierarchical variational conditioning module with an utterance-level speaker encoder to better balance stable timbre preservation and context-adaptive expression. This enables timbre reuse while keeping expression adaptive to the current utterance and, in dialogue, the surrounding context. We also build a comprehensive evaluation protocol for both single-utterance and dialogue settings. Experiments show that CapTalk achieves state-of-the-art performance on a single-utterance voice design benchmark and delivers better expression controllability and contextual appropriateness in multi-turn dialogue. Audio samples are available at: https://anonymous.4open.science/api/repo/Captalk-D601/file/index.html.
\end{abstract}






\maketitle

\section{Introduction}
Recent advances in large-scale multimodal foundation models have reshaped the interaction paradigm of generative artificial intelligence. In image generation, video generation, and other cross-modal generation tasks, natural language has become the most flexible control interface\cite{gemmateam2025gemma3technicalreport,klingteam2026klingmotioncontroltechnicalreport,seedance2025seedance15pronative}. A similar trend is now emerging in speech generation, where voice design aims to synthesize a target voice directly from natural language descriptions, rather than merely reproducing an existing speaker. In this setting, the goal is not only to generate natural and intelligible speech, but also to control who is speaking and how they speak\cite{hu2026voicesculptorvoicedesigned,hu2026qwen3ttstechnicalreport,liu2023promptstyle}. For example, a user may describe a target voice as “a calm middle-aged man with a low and steady voice,” “an energetic young woman speaking brightly and quickly,” or “a police officer loudly scolding a criminal.”

However, compared with visual generation, natural language-driven voice design has emerged much later. Although modern neural TTS systems have achieved substantial progress in naturalness, intelligibility, and zero-shot voice cloning, they still lack a unified and flexible natural language interface for controlling fine-grained attributes such as age, timbre, style, emotion, and character identity\cite{wang2023valle,chen2024valle2,du2024cosyvoice,du2024cosyvoice2scalablestreaming,anastassiou2024seedtts,zhou2025indextts2breakthroughemotionallyexpressive,xie2025controllablespeechsynthesisera,chen2025f5ttsfairytalerfakesfluent}. Early works such as PromptTTS\cite{guo2022promptttscontrollabletexttospeechtext}, PromptTTS2\cite{leng2023prompttts2describinggenerating}, InstructTTS\cite{yang2023instructttsmodellingexpressivetts}, and VoxInstruct\cite{Zhou_2024} have demonstrated the feasibility of using textual descriptions as a control interface for speech, but they still mainly focus on label-based control and do not fully support open-ended voice design with explicit modeling of speaker identity, timbre, and speaking style.

The recent surge of voice design research has been driven by progress in speech understanding. In earlier controllable speech synthesis, systems mainly relied on explicit labels, templated prompts, or implicit style vectors\cite{wang2018gst}. A major bottleneck was the lack of sufficiently strong speech understanding models that could reliably translate complex speaker attributes, timbre characteristics, and expressive styles into high-quality natural language descriptions, making it difficult to construct large-scale language-driven training data. With the development of models such as Qwen3-Omni, Qwen3-Omni-Captioner\cite{xu2025qwen3omnitechnicalreport}, Kimi-Audio\cite{kimiteam2025kimiaudiotechnicalreport}, Step-Audio 2\cite{wu2025stepaudio2technicalreport}, and MiMo-Audio\cite{coreteam2025mimoaudioaudiolanguagemodels}, extracting speaker attributes, style descriptions, and expressive states from audio has become much more feasible. This progress has directly enabled a series of recent voice design works, including FlexiVoice\cite{chen2026flexivoiceenablingflexiblestyle}, VoiceSculptor\cite{hu2026voicesculptorvoicedesigned}, Qwen3TTS-12Hz-1.7B-VD\cite{hu2026qwen3ttstechnicalreport}, MiMo-V2-TTS\cite{coreteam2025mimoaudioaudiolanguagemodels}, Ming-omni-tts-0.5B\cite{inclusionai2026mingomnitts}, and Fish Speech S2 Pro\cite{liao2026fishaudios2}. Despite their diversity, these works still mainly focus on single-utterance generation, and the naturalness of generated speech in voice design settings remains an open challenge. Moreover, for multi-turn dialogue, which is closer to real-world interaction, systematic study is still limited.

Unlike single-utterance generation, dialogue speech generation must not only preserve speaker identity and timbre consistency, but also dynamically adjust turn-level expression and prosodic realization according to the dialogue context. This requires modeling both stable speaker timbre and turn-level expressive variation: the former is better characterized by speaker-level global conditioning, while the latter needs to be explicitly modeled at the current turn\cite{xie2025fireredtts2longconversationalspeech,nguyen2023dgslm,defossez2024moshi}. A central open problem, however, is how to preserve a satisfactorily designed timbre for ongoing generation. Once a desired voice has been crafted through text-guided design,  
  it must be reliably reused across diverse utterances and     
  dialogue contexts. VoiceSculptor addresses this through a    
  two-stage "voice design + cloning" pipeline, in which the 
  designed voice is first rendered into a reference utterance  
  and then reproduced via speaker cloning. This cloning-based  
  strategy, however, introduces a fundamental entanglement:
  while reference audio provides stable timbre cues, it
  inevitably carries utterance-specific emotion, tone, and
  prosody, causing the cloned output to conflate the target
  timbre with transient expressive characteristics of the
  reference. Consequently, achieving both stable timbre
  preservation and context-adaptive expressive control within a
   unified generation framework remains an open challenge.

Beyond model design, evaluation is another major bottleneck for this line of research. Existing public benchmarks provide an important starting point for single-utterance voice design, but they mainly focus on single-utterance speech and often rely on large models to automatically judge consistency between generated speech and textual descriptions\cite{huang2025instructttsevalbenchmarkingcomplexnaturallanguage}. However, natural conversational speech is equally important for real human-computer interaction scenarios. When the task shifts to more natural and less exaggerated speech, as well as context-driven multi-turn dialogue generation, the coverage of existing benchmarks and their agreement with human perception remain limited\cite{feng2025voxprofilespeechfoundationmodel}. We further observe that automatic evaluation is sensitive to audio style distribution. Therefore, while existing benchmarks remain useful references, they should be complemented by evaluation protocols better suited to natural conversational speech and multi-turn dialogue\cite{liao2026fishaudios2,lin2026hearimeanquantifying}.

To address these issues, we propose CapTalk, a unified caption-conditioned text-audio autoregressive framework for both single-utterance and dialogue voice design. The main contributions of this work are summarized as follows:

1. We propose CapTalk, a unified caption-conditioned          
  text-audio autoregressive framework that supports both       
  single-utterance and dialogue voice design under the same    
  discrete token modeling paradigm. In particular,             
  utterance-level captions are used for single-utterance voice 
  design, while speaker-level captions are used to model stable
   speaker timbre and long-term speaking traits in dialogue,   
  and a CoT control sequence is further introduced for
  turn-level expressive control.

2. To address the open problem of preserving a satisfactorily designed timbre while keeping generated expression adaptive to the current utterance and dialogue context, we introduce a Factorized Hierarchical Variational Autoencoder (FHVAE)-inspired hierarchical variational conditioning mechanism that consists of an utterance-level speaker encoder. Specifically, an utterance-level speaker embedding defines the prior of a pooled segment latent, and KL regularization between the segment posterior and this utterance-conditioned prior reinforces stable timbre and voice attributes while attenuating segment-specific affective variation. This allows a desired timbre to be preserved and reused across utterances while keeping generated expression aligned with the current text semantics and, in dialogue, the surrounding context within a unified generation framework.

3. To address the limited coverage of existing voice design evaluation under natural conversational speech and multi-turn dialogue, we build a comprehensive evaluation protocol for both single-utterance and dialogue settings, including single-utterance human evaluation as well as dialogue-oriented CoT prediction, controllability, and comparative evaluation.
\begin{figure*}[t]
  \includegraphics[width=\textwidth]{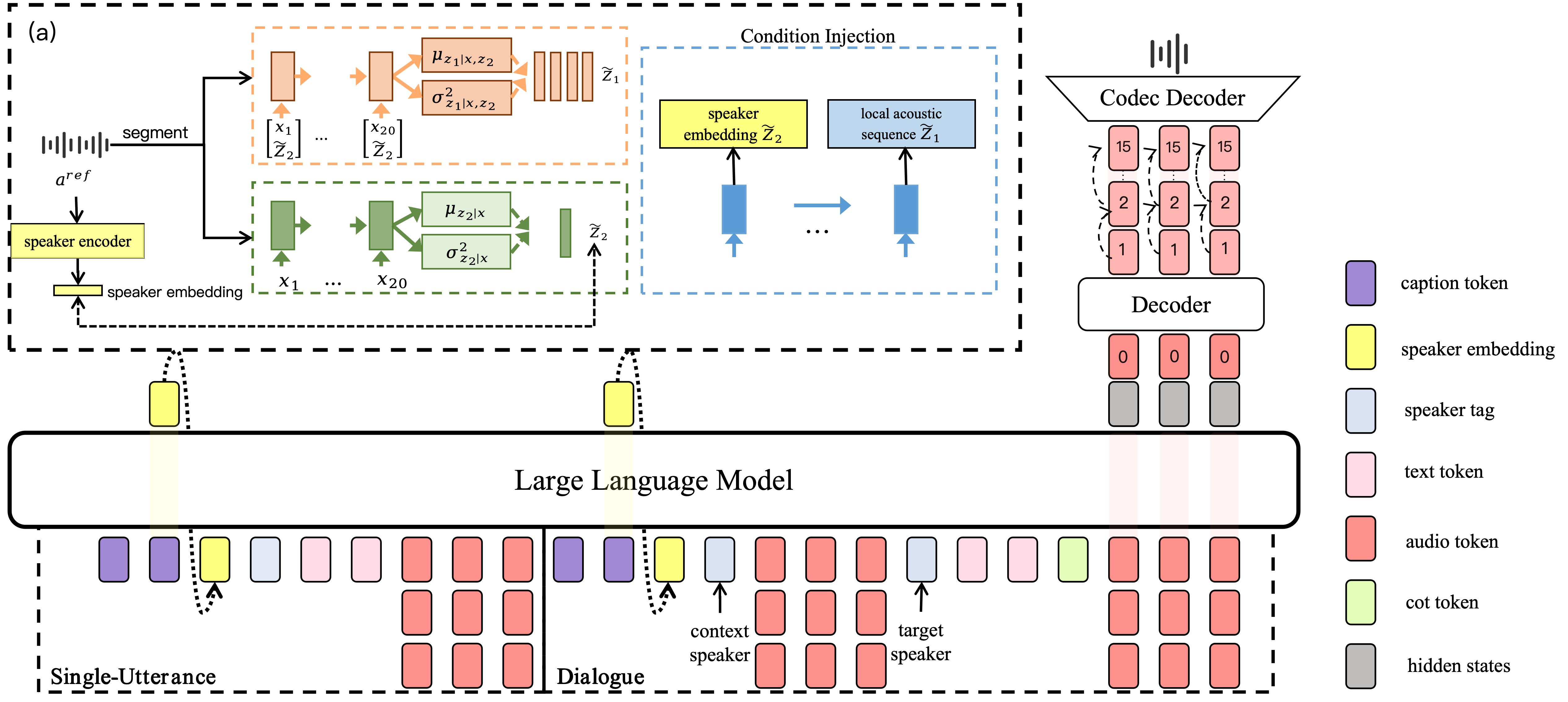}
  \caption{Overview of CapTalk. (a) Hierarchical variational timbre conditioning. The bottom part shows the unified caption-conditioned text-audio autoregressive framework for both single-utterance and dialogue voice design.}
  \Description{Overview of CapTalk with hierarchical variational timbre conditioning and a unified caption-conditioned text-audio autoregressive generation framework.}
  \label{fig:arch}
\end{figure*}
\section{Related Work}

Voice design has recently emerged as a new direction in text-to-speech research, aiming to control speech generation from natural language descriptions rather than predefined style labels or reference audio. Prior work has explored how textual prompts can specify timbre, speaking style, and expressive attributes, enabling more flexible control over speech synthesis\cite{guo2022promptttscontrollabletexttospeechtext,leng2023prompttts2describinggenerating,yang2023instructttsmodellingexpressivetts,Zhou_2024,liu2023promptstyle}. Earlier expressive TTS systems typically relied on latent style embeddings or global style tokens instead of natural-language descriptions\cite{wang2018gst}. More recent systems further combine language-based control with zero-shot TTS or voice cloning. For example, FlexiVoice decouples natural language instructions from speech references, while VoiceSculptor adopts a two-stage design-then-clone pipeline\cite{chen2026flexivoiceenablingflexiblestyle,hu2026voicesculptorvoicedesigned}. Despite their effectiveness, these approaches mainly focus on single-utterance generation.

Beyond these voice-design-focused systems, related developments also appear in broader controllable TTS and unified audio generation frameworks. Qwen3TTS-12Hz-1.7B-VD supports both short-reference voice cloning and natural language-based voice design, while MiMo-V2-TTS extends natural language style control to more open expressive scenarios\cite{hu2026qwen3ttstechnicalreport,coreteam2025mimoaudioaudiolanguagemodels}. Related ideas also appear in unified audio generation systems: Ming-omni-tts-0.5B models speech, music, and sound within a shared framework, and Fish Speech S2 Pro supports multi-speaker and multi-turn generation with instruction-following control\cite{inclusionai2026mingomnitts,liao2026fishaudios2}. In parallel, factorized hierarchical variational models such as FHVAE explicitly separate stable factors from variations, providing useful inspiration for disentangling stable speaker identity and timbre from local acoustic realization\cite{hsu2018scalablefactorizedhierarchicalvariational,Xie_2021}. 

\section{Architecture}

\subsection{Overview}
As shown in Fig.~\ref{fig:arch}, CapTalk is a unified caption-conditioned text-audio autoregressive framework for both single-utterance and dialogue generation. It consists of a speech tokenizer, a dual-transformer text-to-speech model, and a hierarchical variational timbre conditioning module consists of an utterance-level speaker encoder and segment-level latent encoders. The tokenizer converts 16\,kHz speech into discrete acoustic tokens at 12.5\,Hz with 16 codebooks, following recent codec-based speech generation systems\cite{zeghidour2022soundstream,defossez2023encodec}. The text-to-speech model processes captions, an utterance-level speaker  embedding, speaker tags, text, and, when available, dialogue context and CoT control signals in a unified text-audio autoregressive backbone.\cite{wei2023chainofthoughtpromptingelicitsreasoning,hu2025chainofthoughtpromptingspeechtranslation,ma2025audiocotexploringchainofthoughtreasoning}.

The text-to-speech model consists of a backbone Transformer and a lightweight decoder Transformer, where both Transformers are based on the Qwen architecture\cite{qwen2025qwen25technicalreport}. The backbone Transformer models the unified text-speech interleaved sequence and autoregressively predicts the first codebook of each acoustic frame, while the smaller decoder Transformer predicts the remaining codebooks conditioned on the backbone hidden states, the generated first-layer codebook, and the acoustic-side latent conditions $z_2$ and $z_1$. Let $a_t=\{a_t^1,\ldots,a_t^K\}$ denote the discrete acoustic tokens of the $t$-th frame, where $K=16$ is the number of codebooks, and let $x_{<t}$ denote the unified text-audio prefix sequence before frame $t$. We factorize the frame-level probability as
\begin{equation}
p(a_t)=p(a_t^1 \mid x_{<t}) \prod_{k=2}^{K} p(a_t^k \mid x_{<t}, a_t^1, \ldots, a_t^{k-1}).
\end{equation}
The overall model has approximately 1.5B parameters, where the backbone Transformer accounts for the majority and the decoder Transformer, speaker encoder, and hierarchical variational encoders are comparatively lightweight.

\subsection{Single-Utterance and Dialogue Modeling}
CapTalk uses a unified caption-conditioned text-audio autoregressive backbone for both single-utterance and dialogue generation, while incorporating hierarchical variational timbre conditioning into both settings. Concretely, an utterance-level speaker condition $e_{\mathrm{spk}}$ is injected into the backbone to provide stable global speaker information.

\paragraph{Single-Utterance Modeling}
For the single-utterance setting, the backbone input sequence is
\begin{equation}
X^{\mathrm{single}}_{\mathrm{LLM}} = [c_{\mathrm{cap}}, e_{\mathrm{spk}}, c_{\mathrm{txt}}, a].
\end{equation}
Here, $X^{\mathrm{single}}_{\mathrm{LLM}}$ denotes the backbone input sequence for the single-utterance branch, $c_{\mathrm{cap}}$ denotes the caption token subsequence, $e_{\mathrm{spk}}$ denotes the utterance-level speaker  embedding injected into the backbone, $c_{\mathrm{txt}}$ denotes the target text token subsequence with a speaker tag, and $a$ denotes the multi-codebook acoustic token sequence of the target speech. In this setting, $c_{\mathrm{cap}}$ is instantiated from utterance-level captions.

Accordingly, the base training objective of the single-utterance branch is
\begin{equation}
L^{\mathrm{base}}_{\mathrm{single}}
=
2\big((1-\alpha)L_{c0}+\alpha L_{\mathrm{dec}}\big)
+
0.01L_{\mathrm{txt}}.
\end{equation}
Here, $\alpha \in [0,1]$ controls the trade-off between the first-codebook loss and the decoder loss, $L_{c0}$ denotes the cross-entropy loss for predicting the first codebook over all target acoustic frames, $L_{\mathrm{dec}}$ denotes the cross-entropy loss for predicting the remaining codebooks on a randomly sampled subset of target acoustic frames, and $L_{\mathrm{txt}}$ denotes the cross-entropy loss on the supervised target text tokens.

\paragraph{Dialogue Modeling}
For the dialogue setting, the model aims to generate response speech that is consistent with both the target speaker identity and the turn-level speaking style, given the target speaker caption, dialogue history, and the current target text. To this end, the dialogue branch further introduces CoT tokens for explicit control of turn-level dynamic attributes, including emotion, tone, pitch, energy, and speaking rate. The backbone input sequence is
\begin{equation}
X^{\mathrm{dialogue}}_{\mathrm{LLM}} = [c_{\mathrm{cap}}, e_{\mathrm{spk}}, c_{\mathrm{ctx}}, c_{\mathrm{txt}}, c_{\mathrm{cot}}, a].
\end{equation}
Here, $X^{\mathrm{dialogue}}_{\mathrm{LLM}}$ denotes the backbone input sequence for the dialogue branch, $c_{\mathrm{cap}}$ denotes the target speaker caption token subsequence, $e_{\mathrm{spk}}$ denotes the utterance-level speaker  embedding, $c_{\mathrm{ctx}}$ denotes the dialogue-context subsequence from the contexts, $c_{\mathrm{txt}}$ denotes the current target-text subsequence with the target speaker tag, $c_{\mathrm{cot}}$ denotes the CoT token subsequence, and $a$ denotes the target acoustic token sequence. In this setting, $c_{\mathrm{cap}}$ is instantiated from speaker-level captions.

Accordingly, the base training objective of the dialogue branch is
\begin{equation}
L^{\mathrm{base}}_{\mathrm{dialogue}}
=
2\big((1-\alpha)L_{c0}+\alpha L_{\mathrm{dec}}\big)
+
0.01L_{\mathrm{txt}}
+
2.0L_{\mathrm{cot}}.
\end{equation}
Here, $L_{c0}$ and $L_{\mathrm{dec}}$ are defined in the same way as in the single-utterance branch, $L_{\mathrm{txt}}$ denotes the cross-entropy loss on the supervised target text tokens, and $L_{\mathrm{cot}}$ denotes the cross-entropy loss on the CoT tokens.

\paragraph{Overall Objective}
For both settings, the hierarchical variational timbre conditioning module introduces latent prediction losses and KL regularization terms. The final training objective is
\begin{equation}
L
=
L_{\mathrm{base}}
+
\lambda_{\mathrm{spk}} L_{\mathrm{spk\mbox{-}lat}}
+
\lambda_{\mathrm{rec}} L_{\mathrm{rec}}
+
\lambda_{\mathrm{KL}}^{z_2} L_{\mathrm{KL}}^{z_2}
+
\lambda_{\mathrm{KL}}^{z_1} L_{\mathrm{KL}}^{z_1},
\end{equation}
where $L_{\mathrm{base}}$ denotes either $L^{\mathrm{base}}_{\mathrm{single}}$ or $L^{\mathrm{base}}_{\mathrm{dialogue}}$ depending on the training branch, $L_{\mathrm{spk\mbox{-}lat}}$ denotes the latent prediction loss for $e_{\mathrm{spk}}$, $L_{\mathrm{rec}}$ denotes the latent reconstruction loss for $z_2$ and $z_1$, $L_{\mathrm{KL}}^{z_2}$ and $L_{\mathrm{KL}}^{z_1}$ denote the KL regularization terms for $z_2$ and $z_1$, where $z_2$ and $z_1$ are hierarchical variational timbre conditioning module posteriors, and $\lambda_{\mathrm{spk}}$, $\lambda_{\mathrm{rec}}$, $\lambda_{\mathrm{KL}}^{z_2}$, and $\lambda_{\mathrm{KL}}^{z_1}$ are scalar loss weights.

\begin{figure*}[t]
  \includegraphics[width=\textwidth]{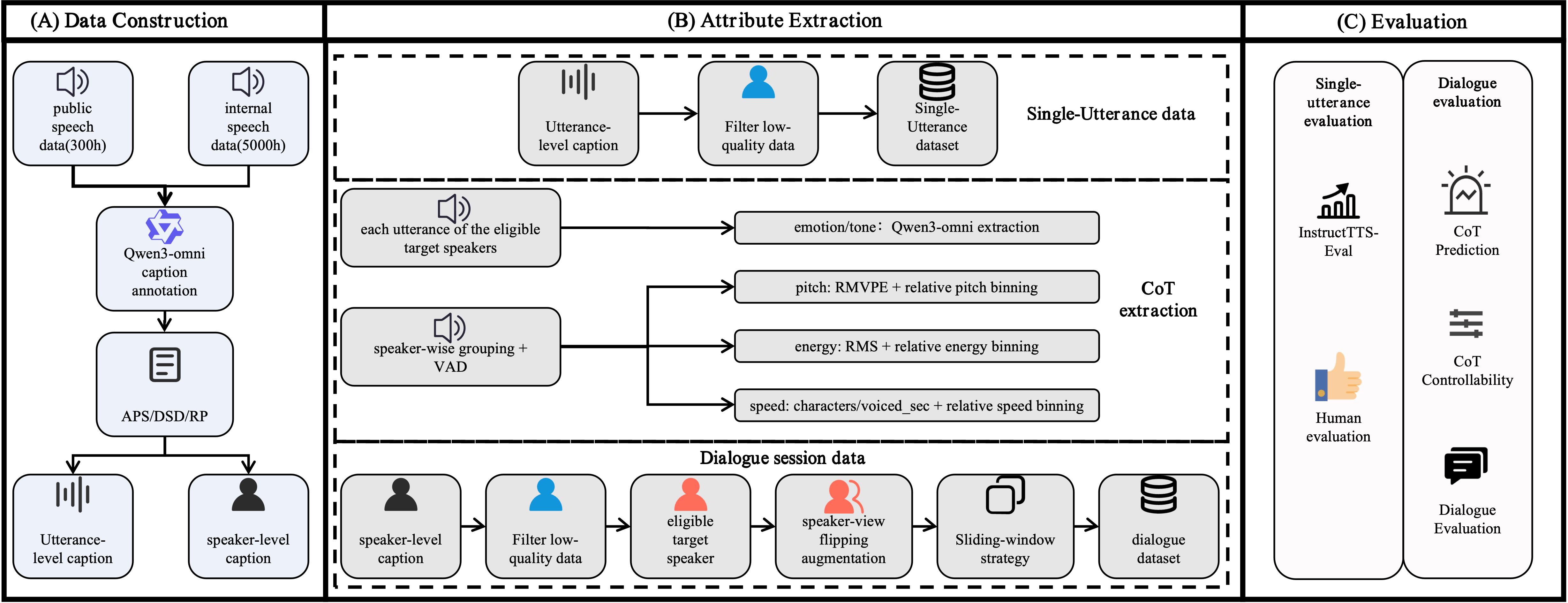}
  \caption{Overview of data construction and evaluation. (A) Qwen3-Omni-based caption annotation for public and internal speech data. (B) Single-utterance and dialogue data construction with dialogue CoT extraction. (C) Evaluation for single-utterance and dialogue voice design.}
  \Description{data_processing}
  \label{fig:data}
\end{figure*}

\subsection{Factorized Hierarchical Variational Timbre Conditioning}
To reduce the entanglement between stable timbre and utterance-specific expression in reference speech, we introduce a hierarchical variational conditioning module inspired by the core factorization idea of FHVAE. Our design is based on two temporal assumptions. First, within a full utterance, speaker timbre and voice attributes such as gender, age, and vocal texture remain approximately constant, whereas emotion and tone may vary over time. Second, within a short segment, emotion and tone can be treated as locally stationary. Under these assumptions, a pooled full-utterance representation preserves stable speaker factors while smoothing time-varying affective factors, whereas a pooled segment representation retains the same stable speaker factors together with a segment-local snapshot of emotion and tone.

Accordingly, we use two granularities of latent conditioning. A speaker encoder with global temporal pooling extracts an utterance-level speaker  embedding $e_{\mathrm{spk}}$ from the full reference utterance. In parallel, short segments are encoded into a pooled latent $z_2$ and a frame-level latent $z_1$. Because $z_2$ is obtained from pooled segment representations, it tends to capture factors that are locally constant within the segment, including timbre, speaker attributes, and the local affective state. By contrast, $z_1$ captures finer-grained frame-level variation such as phonetic content and prosody. The key idea is then to regularize the segment posterior $q(z_2 \mid s)$ toward an utterance-conditioned prior $p(z_2 \mid e_{\mathrm{spk}})$. Since $e_{\mathrm{spk}}$ and $z_2$ share stable speaker  information but do not represent emotion and tone at the same temporal granularity, this KL regularization encourages the shared stable component to be preserved while attenuating segment-specific affective variation in $z_2$.

Let $\mathcal{S}(a^{\mathrm{ref}})=\{s^{(m)}\}_{m=1}^{M}$ denote the set of short segments extracted from a reference speech sample $a^{\mathrm{ref}}$. We first extract the utterance-level speaker  embedding from the full utterance:
\begin{equation}
e_{\mathrm{spk}} = \mathrm{SpeakerEnc}(a^{\mathrm{ref}}),
\end{equation}
where $\mathrm{SpeakerEnc}(\cdot)$ denotes a speaker encoder with global temporal pooling.

For notational simplicity, we write the hierarchical posterior for a generic segment $s \in \mathcal{S}(a^{\mathrm{ref}})$ and average the corresponding losses over all segments from the same reference utterance. The hierarchical posterior is
\begin{equation}
q_{\phi}(z_1, z_2 \mid s)
=
q_{\phi_{2}}(z_2 \mid s)
\cdot
q_{\phi_{1}}(z_1 \mid s, z_2),
\end{equation}
where $q_{\phi_{2}}(z_2 \mid s)$ is the posterior of the pooled segment latent and $q_{\phi_{1}}(z_1 \mid s, z_2)$ is the posterior of the frame-level latent conditioned on both the segment and $z_2$. Here, $z_2 \in \mathbb{R}^{d_2}$ denotes the pooled segment latent, and $z_1$ denotes the frame-level latent used in the acoustic generation pathway. Their Gaussian posteriors are defined as
\begin{equation}
q_{\phi_{2}}(z_2 \mid s)
=
\mathcal{N}\!\big(
g_{\mu_{z_2}}(s),\;
\operatorname{diag}(g_{\sigma_{z_2}}^2(s))
\big),
\end{equation}
\begin{equation}
q_{\phi_{1}}(z_1 \mid s, z_2)
=
\mathcal{N}\!\big(
g_{\mu_{z_1}}(s, z_2),\;
\operatorname{diag}(g_{\sigma_{z_1}}^2(s, z_2))
\big),
\end{equation}
where $g_{\mu_{z_2}}(\cdot)$ and $g_{\sigma_{z_2}}(\cdot)$ are the mean and scale predictors for $z_2$, and $g_{\mu_{z_1}}(\cdot,\cdot)$ and $g_{\sigma_{z_1}}(\cdot,\cdot)$ are the mean and scale predictors for $z_1$.

We use a standard normal prior for $z_1$ and an utterance-conditioned prior for $z_2$:
\begin{equation}
\mu_{\mathrm{spk}} = f_{\eta}(e_{\mathrm{spk}}),
\end{equation}
\begin{equation}
p(z_2 \mid e_{\mathrm{spk}})=\mathcal{N}(\mu_{\mathrm{spk}}, I),
\qquad
p(z_1)=\mathcal{N}(0,I),
\end{equation}
where $f_{\eta}(\cdot)$ projects the utterance-level speaker  embedding to the mean of the prior distribution for $z_2$, and $I$ is the identity covariance matrix. 

The corresponding KL regularizers are
\begin{equation}
L_{\mathrm{KL}}^{z_2}
=
\frac{1}{M}\sum_{m=1}^{M}
D_{\mathrm{KL}}
\!\left(
q_{\phi_{2}}(z_2 \mid s^{(m)})
\;\|\;
p(z_2 \mid e_{\mathrm{spk}})
\right),
\end{equation}
\begin{equation}
L_{\mathrm{KL}}^{z_1}
=
\frac{1}{M}\sum_{m=1}^{M}
D_{\mathrm{KL}}
\!\left(
q_{\phi_{1}}(z_1 \mid s^{(m)}, z_2)
\;\|\;
\mathcal{N}(0,I)
\right).
\end{equation}
The term $L_{\mathrm{KL}}^{z_2}$ encourages the pooled segment latent $z_2$ to remain close to the utterance-level speaker  representation whenever the two agree, thereby reinforcing stable timbre-related information while discouraging segment-specific variation that is not consistent across the full utterance.

To enable inference without reference speech, we further predict the utterance-level from textual conditions:
\begin{equation}
\hat{e}_{\mathrm{spk}} = f_{\psi}(h_{\mathrm{cap}}), \qquad
\end{equation}
where $f_{\psi}(\cdot)$ is prediction head, $h_{\mathrm{cap}}$ denotes the last hidden states of the caption segment. We align them with the corresponding training-time representations by
\begin{equation}
L_{\mathrm{spk-lat}} = \left\| \hat{e}_{\mathrm{spk}} - \operatorname{sg}(e_{\mathrm{spk}}) \right\|_2^2,
\end{equation}
where $\operatorname{sg}(\cdot)$ denotes the stop-gradient operator and $\|\cdot\|_2^2$ denotes the squared Euclidean distance.

At training time, $e_{\mathrm{spk}}$ is injected into the backbone Transformer as a stable timbre condition. At inference time, the model supports both caption-guided generation, where $\hat{e}_{\mathrm{spk}}$ is predicted from textual conditions, and timbre-reuse generation, where a previously obtained $e_{\mathrm{spk}}$ is fixed to preserve the designed timbre.

\section{Task Formulation, Data Construction, and Evaluation Protocol}

This section introduces the data construction, attribute extraction, and evaluation protocol for both single-utterance and multi-turn dialogue voice design. The overall pipeline is illustrated in Fig.~\ref{fig:data}. We first describe the data sources and caption construction; we then introduce the CoT design and attribute extraction procedure for dialogue; finally, we build a comprehensive evaluation protocol for both single-utterance and dialogue generation.
\subsection{Data Sources and Caption Construction}
We use both public and internal datasets. The single-utterance experiments are conducted on 300 hours of public data and 5000 hours of internal data. The dialogue experiments use full two-speaker dialogue sessions from a subset of the internal speech corpus. Overall, the data cover both acted-style speech and natural conversational speech.

In the caption annotation stage, we use Qwen3-Omni, a multimodal model, to generate natural-language descriptions and construct two types of captions: utterance-level captions for individual speech samples, which describe utterance-level expressive characteristics, and speaker-level captions for speaker-aggregated speech, which characterize relatively stable speaker attributes and long-term speaking traits. For speaker-level captions, we sample ten speech segments from the same speaker and jointly input them into Qwen3-Omni to generate the description. In the speaker-level description design, we still retain dynamic attributes such as pitch variations and volume variations.

To unify speaker description formats, we follow the three description styles in InstructTTSEval, namely Acoustic-Parameter Specification (APS), Descriptive-Style Directive (DSD), and Role-Play (RP). In InstructTTSEval, APS consists of explicit instructions covering twelve features: gender, pitch, speaking rate, volume, age, clarity, fluency, accent, timbre texture, emotion, tone, and personality. DSD rewrites such structured instructions into free-form natural-language descriptions, while RP describes roles or scenarios that imply a target speaking style. For both the single-utterance and dialogue settings, we generate all three caption styles for each sample, namely APS, DSD, and RP. During training, these three caption variants are all used as conditioning inputs, so that each underlying sample can contribute multiple caption-conditioned training instances.

For the single-utterance setting, we use only utterance-level captions as conditioning inputs. After caption construction, we further filter out low-quality samples and retain the remaining utterance-caption pairs to form the final single-utterance dataset.

For the dialogue setting, we use only speaker-level        
  captions as speaker conditions. Dialogue data are organized with  
  full two-speaker sessions as the basic units. Starting from  
  the speaker-level captions, we further filter low-quality    
  data and determine the eligible target speakers in each      
  session. Here, an eligible target speaker refers to a speaker
   whose utterances are of sufficient quality to be used as
  supervised target speech in training. If
  only one speaker in a dialogue session passes the quality
  filter, that speaker is treated as the only eligible target
  speaker, while the other speaker is used only as the context
  speaker. If both speakers pass the quality filter, both are
  treated as eligible target speakers and contribute training
  samples from two speaker views: in one view, speaker A is
  selected as the target speaker and speaker B provides the
  dialogue context, while in the other view, speaker B is
  selected as the target speaker and speaker A provides the
  dialogue context. If neither speaker passes the quality
  filter, the dialogue session is discarded. Based on the
  eligible target speakers, we then convert each dialogue
  session into training samples. In addition, because many
  dialogue sessions are too long to fit into a single training
  sequence, we apply a sliding-window strategy over dialogue
  turns. Specifically, we use a window size of 20 turns and
  split each long dialogue session into multiple dialogue
  segments, which together form the final dialogue dataset.

\subsection{CoT Design and Attribute Extraction}
After determining the eligible target speakers, we extract CoT-style attributes from each target utterance used for supervision. The current CoT consists of five attributes: emotion, tone, pitch, energy, and speaking rate, arranged in the fixed order <emotion><tone><pitch> <energy><speed>. Among them, emotion and tone describe high-level turn-wise expressive states, while pitch, energy, and speaking rate characterize lower-level prosodic realization. We model the latter as speaker-internal relative prosody rather than absolute comparisons across speakers.

The choice of these five attributes follows a hierarchical view of dialogue expression. Emotion and tone act as high-level expressive factors that reflect the speaker's affective state and communicative intent at the current turn\cite{ma2023emotion2vecselfsupervisedpretrainingspeech,gao2024emodpocontrollableemotionalspeech,zhou2021seenunseenemotionalstyle}. In turn, they are typically realized through lower-level prosodic variation, which we model using pitch, energy, and speaking rate\cite{raitio2020controllableneuraltexttospeechsynthesis,ladd2008intonational}. This design yields a compact CoT representation that links high-level dialogue expression to its lower-level acoustic realization.

In practice, emotion and tone are extracted from each utterance of the eligible target speakers by Qwen3-Omni. For pitch, energy, and speaking rate, we first group speech by speaker and apply VAD to retain only speech-active segments, so that the extracted attributes can be normalized relative to each speaker's own speaking baseline. For pitch, we extract voiced-frame F0 trajectories with RMVPE\cite{Wei2023RMVPE}, use the median F0 as the utterance-level pitch value, and map it into a semitone space with 440\,Hz as reference. A speaker-specific median semitone value is then used as the baseline to derive relative pitch deviation, which is discretized into normal, slightly high/low, noticeably high/low, and extremely high/low. For energy, RMS is computed on the voiced-only waveform and normalized by the speaker-level median RMS, then discretized into normal, slightly louder/quieter, noticeably louder/quieter, and extremely louder/quieter. For speaking rate, punctuation, symbols, and spaces are removed from the text, and a character-level rate is computed as ``number of characters / voiced\_sec''. This is then normalized by the speaker-level median speaking rate and discretized into normal, slightly faster/slower, noticeably faster/slower, and extremely faster/slower. Therefore, pitch, energy, and speaking rate all characterize deviations relative to each speaker's normal speaking state rather than absolute attributes across speakers.
\begin{table}[t]
  \caption{Results on the InstructTTSEval-ZH Benchmark.}
  \label{tab:instructttseval-zh}
  \begin{tabular}{lcccc}
    \toprule
    Model & APS & DSD & RP & AVG \\
    \midrule
    Qwen3TTS-12Hz-1.7B-VD & \textbf{87.10} & \textbf{76.00} & \underline{55.20} & \underline{72.77} \\
    Ming-omni-tts-0.5B & \underline{84.90} & 72.20 & 53.90 & 70.33 \\
    VoiceSculptor & 73.77 & 65.40 & 47.60 & 62.26 \\
    Fish Speech S2 Pro & 29.61 & 50.80 & 42.60 & 41.00 \\
    CapTalk-1.5B (Ours) & 84.10 & \underline{75.40} & \textbf{61.70} & \textbf{73.73} \\
    \bottomrule
  \end{tabular}
\end{table}
\subsection{Evaluation Protocol and Benchmark Analysis}
Evaluation for voice design is still at an early stage. We first adopt InstructTTSEval as a public benchmark for single-utterance controllable speech evaluation, but its setting does not fully match the target scenarios considered in this work. First, the benchmark mainly relies on large models to automatically judge the consistency between generated speech and textual descriptions, without sufficient human validation. Second, it mainly focuses on single-utterance speech and does not cover context dependence, target speaker consistency, or turn-level style control in multi-turn dialogue. In addition, natural conversational speech is typically flatter and more restrained, so relying solely on existing automatic benchmarks may not fully reflect model performance in this setting. Therefore, we treat the existing benchmark as a starting point for single-utterance automatic evaluation, and further complement it with human evaluation and a dialogue-specific evaluation protocol.

For the single-utterance task, we retain InstructTTSEval as an automatic evaluation reference and additionally conduct human subjective evaluation to compensate for its limited coverage in natural conversational speech. Specifically, we score generated speech on a 1--5 scale from seven aspects: overall description consistency, identity consistency, timbre consistency, expressive-style consistency, role-intent consistency, control stability, and MOS naturalness.  Detailed definitions of these evaluation aspects are provided in Appendix~B.1.

For the dialogue task, since no existing benchmark provides a corresponding setting, we further propose an extended evaluation protocol with three components. First, we evaluate the plausibility of CoT prediction for the current turn. Since multiple CoT combinations may be reasonable under the same context, this task is not suitable for exact matching against a unique ground-truth answer. Instead, we combine Gemini-based evaluation and human evaluation to assess whether the predicted CoT is consistent with contextual semantics, role state, and turn-level expressive needs, and report the corresponding accuracy metrics\cite{jia2024leveragingllmsdialoguequality}. Second, we evaluate CoT controllability by changing only one CoT attribute while fixing the text, context, and random seed, and then examining whether the generated speech changes as expected along the target dimension. For emotion and tone, we use Qwen3-Omni for posterior extraction; for pitch, energy, and speaking rate, we use the same relative prosody extraction method as in training for quantitative evaluation. Finally, we conduct overall dialogue evaluation to analyze the practical contribution of CoT control to generation quality. Specifically, we train an additional model variant without CoT input and compare it with the full model under the same dialogue context to assess whether CoT provides perceptible gains for dialogue generation.

Overall, we complement existing single-utterance benchmarks with human evaluation and extend evaluation to dialogue-specific CoT prediction, CoT controllability, and overall dialogue quality, thereby forming a more suitable evaluation framework for our task setting.
\begin{table*}[t]
  \caption{Human evaluation results for single-utterance voice design. All metrics are rated on a 1--5 scale. MOS denotes human-rated naturalness. Best results are in \textbf{bold}.}
  \label{tab:single-human-eval}
  \centering
  \setlength{\tabcolsep}{5pt}
  \begin{tabular}{@{}lccccccc@{}}
    \toprule
    Model & Overall & Identity & Timbre & Express. & Role & Stabil. & MOS \\
    \midrule
    Ming-omni-tts-0.5B & 3.95 & \textbf{4.22} & 4.03 & 4.06 & 3.88 & 4.27 & 3.91 \\
    Qwen3TTS-12Hz-1.7B-VD & 4.20 & 3.78 & \textbf{4.15} & \textbf{4.12} & 3.85 & 4.35 & 3.82 \\
    VoiceSculptor & 3.00 & 3.15 & 2.89 & 2.74 & 2.81 & 3.25 & 2.87 \\
    Fish Speech S2 Pro & 2.07 & 2.19 & 2.02 & 1.82 & 1.86 & 2.29 & 2.11 \\
    \midrule
    CapTalk (Ours) & \textbf{4.24} & 3.98 & 3.59 & 4.10 & \textbf{4.17} & \textbf{4.38} & \textbf{4.20} \\
    \bottomrule
  \end{tabular}
\end{table*}

\begin{table*}[t]
  \caption{Automatic evaluation results for dialogue voice design.}
  \label{tab:dialogue-auto-eval}
  \centering
  \setlength{\tabcolsep}{7pt}
  \begin{tabular}{@{}llccccc@{}}
    \toprule
    Evaluation Category & Metric & Emotion & Tone & Pitch & Energy & Speed \\
    \midrule
    CoT Prediction & Accuracy & 0.7850 & 0.7675 & 0.8375 & 0.8250 & 0.9125 \\
    CoT Controllability & Success Rate & 0.7675 & 0.7675 & 0.8400 & 0.8550 & 0.8675 \\
    \bottomrule
  \end{tabular}
\end{table*}

\begin{table}[t]
  \caption{Context coherence comparison in dialogue evaluation.}
  \label{tab:dialogue-context-coherence}
  \centering
  \begin{tabular*}{\columnwidth}{@{\extracolsep{\fill}}lcc@{}}
    \toprule
    Model Setting & Gemini & Human Pairwise Preference \\
    \midrule
    Dialogue Eval. (w/ CoT) & \textbf{72\%} & \textbf{65.5\%} \\
    Dialogue Eval. (w/o CoT) & 28\% & 34.5\% \\
    \bottomrule
  \end{tabular*}
\end{table}
\section{Experiments}
\subsection{Comparative Experiments}

To verify the effectiveness of our method in both single-utterance and multi-turn dialogue voice design, we systematically compare it with representative voice design and controllable TTS models. Since the single-utterance and dialogue settings differ in both task form and evaluation objective, we evaluate them separately: the single-utterance setting uses a public benchmark and human evaluation, while the dialogue setting adopts an extended evaluation protocol targeting CoT control and context-aware generation.

\subsubsection{Single-Utterance Comparison}
In the single-utterance setting, we first evaluate on the InstructTTSEval-ZH benchmark, which measures controllability from three aspects, namely APS, DSD, and RP, and uses their average (AVG) as the overall metric. We compare our method with Qwen3TTS-12Hz-1.7B-VD, Ming-omni-tts-0.5B, VoiceSculptor, and Fish Speech S2 Pro. As shown in Table~\ref{tab:instructttseval-zh}, existing models perform relatively stably on APS and DSD, but generally show a clear drop on RP, indicating that role-level intent and persona modeling from natural language remain challenging. Our method achieves the best RP score and the highest overall performance, surpassing the second-best Qwen3TTS-12Hz-1.7B-VD. Although Qwen3TTS-12Hz-1.7B-VD leads on APS and DSD, its RP score drops notably, whereas our method achieves more balanced performance across all three dimensions.

Since the public benchmark cannot fully reflect performance in natural conversational speech, we further report human subjective evaluation in Table~\ref{tab:single-human-eval}. As shown in Table~\ref{tab:single-human-eval}, CapTalk achieves the best overall score, role-intent consistency, control stability, and MOS, indicating that it better balances naturalness and high-level instruction following under more conversational conditions. One possible explanation for the relatively lower timbre-consistency score is that our single-utterance training data are dominated by natural conversational speech, whose fine-grained timbre variation is typically less salient than that of acted-style speech. As a result, the model may learn naturalness and high-level role/style intent more effectively than subtle timbre texture under text-only voice design.

\begin{table}[t]
  \caption{Scaling law results for single-utterance voice design on InstructTTSEval-ZH.}
  \label{tab:scaling-single}
  \centering
  \setlength{\tabcolsep}{6pt}
  \begin{tabular}{@{}llcccc@{}}
    \toprule
    Data Scale & Type & APS & DSD & RP & AVG \\
    \midrule
    300h & Acted & 82.78 & 71.41 & 49.10 & 67.76 \\
    1500h & Conversat. & 70.20 & 57.80 & 41.20 & 56.40 \\
    3000h & Conversat. & 74.00 & 62.40 & 46.80 & 61.07 \\
    5000h & Conversat. & 79.10 & 68.30 & 55.10 & 67.50 \\
    5000h+300h & Conversat.+Acted & \textbf{84.10} & \textbf{75.40} & \textbf{61.70} & \textbf{73.73} \\
    \bottomrule
  \end{tabular}
\end{table}

\begin{table*}[t]
  \caption{Scaling law results for dialogue voice design under CoT prediction and CoT controllability evaluation.}
  \label{tab:scaling-dialogue}
  \centering
  \setlength{\tabcolsep}{5pt}
  \begin{tabular}{llccccc}
    \toprule
    Data Scale & Evaluation Category & Emotion & Tone & Pitch & Energy & Speed \\
    \midrule
    17,258 sess., 3373 h & CoT Prediction & 0.7325 & 0.6950 & 0.8050 & 0.7875 & 0.8825 \\
    24,198 sess., 4482.35 h & CoT Prediction & 0.7550 & 0.7225 & 0.8250 & 0.8025 & 0.9000 \\
    32,171 sess., 6270.68 h & CoT Prediction & \textbf{0.7850} & \textbf{0.7675} & \textbf{0.8375} & \textbf{0.8250} & \textbf{0.9125} \\
    \midrule
    17,258 sess., 3373 h & CoT Controllability & 0.7025 & 0.6900 & 0.7800 & 0.8125 & 0.8275 \\
    24,198 sess., 4482.35 h & CoT Controllability & 0.7375 & 0.7300 & 0.8125 & 0.8350 & 0.8475 \\
    32,171 sess., 6270.68 h & CoT Controllability & \textbf{0.7675} & \textbf{0.7675} & \textbf{0.8400} & \textbf{0.8550} & \textbf{0.8675} \\
    \bottomrule
  \end{tabular}
\end{table*}

\subsubsection{Dialogue Comparison}
Following the evaluation protocol described in Section~4.3, we evaluate dialogue voice design from three perspectives. For CoT prediction plausibility, Gemini-assisted and human evaluation on 400 samples show that our model achieves reasonable prediction accuracy across all five attributes (Table~\ref{tab:dialogue-auto-eval}), with pitch, energy, and speed being the most accurate. For CoT controllability, the controlled generation experiments confirm that varying a single CoT attribute produces the expected change in the corresponding acoustic dimension (Table~\ref{tab:dialogue-auto-eval}). For overall dialogue quality, Table~\ref{tab:dialogue-context-coherence} shows that the model with CoT is preferred over the variant without CoT by both Gemini (72\% vs.\ 28\%) and human evaluators (65.5\% vs.\ 34.5\%), confirming that explicit CoT planning improves context coherence in dialogue generation. Detailed dialogue evaluation materials are provided in Appendix~B.2, including the dialogue evaluation protocol, the details of CoT prediction evaluation, and additional human-evaluation details.

\paragraph{Cross-System Dialogue Comparison.}
\begin{table}[t]
\setlength{\tabcolsep}{6pt}
\caption{Cross-system dialogue comparison between CapTalk and Fish Speech S2 Pro. SIM measures cross-turn target-speaker timbre consistency in dialogue; Context Coherence and MOS are human ratings (1--5).}
\label{tab:cross-system}
\centering
\begin{tabular}{lccc}
\toprule
Model & SIM$\uparrow$ & Context Coherence$\uparrow$ & MOS$\uparrow$ \\
\midrule
Fish S2 Pro & 0.806 & 3.61 & 3.80 \\
CapTalk & \textbf{0.808} & \textbf{4.18} & \textbf{4.12} \\
\bottomrule
\end{tabular}
\end{table}

We further compare CapTalk with Fish Speech S2 Pro on 40 multi-turn dialogues (Table~\ref{tab:cross-system}). In this dialogue setting, SIM is used to judge whether multiple generated utterances of the target speaker sound as if they are spoken by the same person, and therefore primarily reflects cross-turn timbre consistency. Concretely, for each dialogue session, we extract speaker embeddings from all generated target-speaker utterances using a pretrained WavLM model\cite{chen2021wavlm} and compute their average pairwise cosine similarity. Higher SIM indicates stronger consistency of the target speaker's timbre across turns. Context coherence and MOS are evaluated through human rating on a 1--5 scale. CapTalk achieves higher scores on all three metrics than Fish Speech S2 Pro, indicating that the hierarchical variational timbre conditioning better preserves target-speaker timbre across dialogue turns, while explicit CoT planning contributes to better context coherence and naturalness.

\subsection{Scaling Law Experiments}
We further study how data scale and style coverage affect voice design in both single-utterance and multi-turn dialogue settings.

\subsubsection{Scaling Law in Single-Utterance Voice Design}
In the single-utterance setting, we train on acted, conversational, and mixed-scale data and evaluate on InstructTTSEval-ZH (Table~\ref{tab:scaling-single}). Performance depends on both data scale and style composition: 1500-hour conversational data underperforms 300-hour acted data, whereas 5000-hour conversational data improves substantially, and adding 300 hours of acted speech yields the best result (AVG 73.73). This suggests that natural conversational data provides broader coverage, while acted speech is useful for stronger expressive attributes.

\subsubsection{Scaling Law in Dialogue Voice Design}
In the dialogue setting, we train on 17,258, 24,198, and 32,171 sessions and report CoT prediction and controllability in Table~\ref{tab:scaling-dialogue}. Performance improves steadily with data scale across all five attributes, indicating that dialogue voice design benefits directly from large-scale contextualized multi-speaker data.

\subsection{Ablation Study}
\begin{table}[t]
  \caption{Ablation study on explicit caption\_loss supervision in the single-utterance setting.}
  \label{tab:ablation-caption-loss}
  \centering
  \setlength{\tabcolsep}{6pt}
  \begin{tabular}{lcccc}
    \toprule
    Type & APS & DSD & RP & AVG \\
    \midrule
    Baseline (w/o caption\_loss) & \textbf{79.10} & 68.30 & \textbf{55.10} & \textbf{67.50} \\
    w/ caption\_loss & 77.20 & \textbf{69.10} & 54.40 & 66.90 \\
    \bottomrule
  \end{tabular}
\end{table}

We study the effect of explicit caption\_loss supervision in the single-utterance setting, while the dialogue no-CoT ablation is reported in the comparative experiments. Table~\ref{tab:ablation-caption-loss} shows that adding caption\_loss slightly improves DSD from 68.30 to 69.10, but reduces APS from 79.10 to 77.20, RP from 55.10 to 54.40, and the overall AVG from 67.50 to 66.90. These results suggest that explicit caption supervision is not beneficial overall in our setting, and we therefore disable caption\_loss in the final model.

\section{Conclusion}
We propose CapTalk, a unified caption-conditioned text-audio autoregressive framework for both single-utterance and dialogue voice design. CapTalk uses caption-based conditioning for voice design, with utterance-level captions in the single-utterance setting and speaker-level captions for target speaker modeling in dialogue, and further introduces a CoT control sequence for turn-level dynamic expression planning. Inspired by the core factorization idea of FHVAE, our hierarchical variational conditioning module consists of an utterance-level speaker encoder, using an utterance-conditioned prior over the pooled segment latent to reinforce stable timbre and voice attributes while suppressing segment-specific affective variation. This enables timbre reuse while keeping expression adaptive to dialogue context. Experiments show that CapTalk achieves state-of-the-art overall performance on single-utterance voice design and delivers better expression controllability, and context coherence in multi-turn dialogue. We also build a comprehensive evaluation protocol covering both settings. We plan to release caption annotations and a subset of data upon acceptance to facilitate future research.


\clearpage
\bibliographystyle{ACM-Reference-Format}
\bibliography{captalk-ref}

\clearpage
\appendix
\section{Limitations and Future Work}
Although our method achieves promising results in both single-utterance and multi-turn dialogue voice design, several limitations remain. First, our data annotation pipeline relies heavily on the audio understanding and caption generation capabilities of Qwen3-Omni, whose description quality still has inherent limitations; this in turn affects the accuracy and stability of caption conditions. Second, our training data mainly consists of natural conversational speech (casual chat style), whose emotional intensity and expressive range are typically weaker than those of acted-style speech, leaving room for improvement in more expressive settings.

We identify three directions for future work. First, we plan to explore multi-model cross-validation and human-in-the-loop correction to improve caption quality and reduce description bias. Second, constructing a larger-scale voice design corpus with broader coverage of age, accent, and expressive style would benefit generalization. Third, building on the evaluation protocol proposed in this work, we plan to develop a high-quality, publicly available benchmark specifically designed for voice design, covering both single-utterance and dialogue scenarios with human-validated annotations, to provide a more reliable evaluation foundation for the community.

\section{Additional Evaluation Details}
\label{sec:appendix-eval}

\subsection{Single-Utterance Human Evaluation}

For single-utterance voice design, human evaluation is conducted from the following seven aspects, each rated on a 1--5 scale.

\paragraph{Overall Description Consistency.}
This metric evaluates the overall consistency between the generated audio and the caption description from a holistic listening perspective.

\paragraph{Identity Consistency.}
This metric evaluates whether the generated audio matches the caption in terms of basic identity-related attributes, including perceived gender, age, and accent/regional characteristics.

\paragraph{Timbre Consistency.}
This metric evaluates whether the generated audio matches the caption in terms of timbre-related properties, including voice quality, timbre texture, brightness, thickness, breathiness, and hoarseness.

\paragraph{Expressive-Style Consistency.}
This metric evaluates whether the generated audio matches the caption in terms of expression, including emotion, prosody, and speaking style, i.e., how the utterance is delivered.

\paragraph{Role-Intent Consistency.}
This metric evaluates whether the generated audio reflects the higher-level role perception, scene perception, and expressive intent described in the caption.

\paragraph{Control Stability.}
This metric evaluates the consistency of generated outputs in terms of vocal characteristics and expressive style across multiple generations under the same caption.

\paragraph{MOS (Naturalness).}
This metric evaluates whether the generated audio sounds natural and human-like, regardless of the caption condition. Annotators rate each sample on a 1--5 scale, and the final MOS is obtained by averaging scores across annotators and samples.

\subsection{Dialogue Evaluation Protocol}

Table~\ref{tab:appendix-dialogue-eval} summarizes the dialogue evaluation protocol.

\begin{table*}[t]
  \caption{Summary of the dialogue evaluation protocol.}
  \label{tab:appendix-dialogue-eval}
  \begin{tabular}{>{\centering\arraybackslash}p{3.0cm}p{4.0cm}p{4.8cm}p{3.2cm}}
    \toprule
    Evaluation Category & Purpose & Method & Reported Results \\
    \midrule
    CoT Prediction Evaluation &
    Evaluate whether the predicted CoT for the current utterance is reasonable &
    Gemini-assisted evaluation and human evaluation are used to judge whether the predicted CoT is consistent with the dialogue context &
    Accuracy \\
    CoT Controllability Evaluation &
    Evaluate the effectiveness of CoT-based control &
    Controlled generation is performed by fixing the text, context, and random seed while changing only one CoT attribute at a time &
    Success Rate \\
    Overall Dialogue Evaluation &
    Evaluate whether introducing CoT brings benefits to dialogue generation &
    Gemini-assisted evaluation and human pairwise evaluation are used to compare dialogue generation with and without CoT &
    Context Coherence Preference \\
    \bottomrule
  \end{tabular}
\end{table*}

\subsubsection{Details of CoT Prediction Evaluation}

For the dialogue setting, we do not treat exact matching between predicted CoT and ground-truth CoT as the only evaluation criterion. Instead, we evaluate whether the predicted CoT is reasonable given the target speaker profile, dialogue history, and target utterance. This design is motivated by the fact that, in dialogue, a single target utterance may admit multiple reasonable stylistic interpretations. Therefore, comparison against a single annotated answer may underestimate the actual CoT prediction ability of the model.

Specifically, for each target turn, the model predicts a CoT representation composed of emotion, tone, pitch, energy, and speed based on the caption, dialogue history, and current text. The predicted CoT is then judged for plausibility with access to the target speaker profile, the most recent four turns of dialogue text and their corresponding audio, the current target text. During evaluation, the judge model is not provided with the ground-truth CoT, so as to avoid direct bias from reference labels.

Evaluation samples are stratified and randomly sampled from the validation set. We select 100 sessions in total. For each session, four target turns are further selected to cover the early, early-middle, late-middle, and late parts of the dialogue, so as to ensure a balanced position distribution within each session. This yields 400 evaluation samples in total. 

For automatic evaluation, Gemini is used as the judge model. For each sample, Gemini gives a binary plausible/implausible judgment for each of the five CoT attributes and also outputs an overall judgment. In the main paper, we primarily report the plausibility rate of the five attribute dimensions, while the overall judgment is used as an auxiliary indicator. In addition, we conduct human evaluation on a smaller subset to verify the reliability of the automatic evaluation.

To validate the reliability of Gemini-based evaluation, we further perform human evaluation. Specifically, 30 samples are stratified and randomly selected from the CoT prediction evaluation set, and 15 annotators independently judge whether the predicted CoT is reasonable. The annotators are given access to the target speaker caption, the most recent four turns of dialogue text and their corresponding audio, the current target text, and the predicted CoT; the ground-truth CoT is not provided.

The evaluation criterion is the same as that used in automatic evaluation, namely, whether the predicted CoT can be regarded as a reasonable interpretation of the current utterance given the dialogue history and speech context. Annotators provide binary judgments for the five CoT attributes as well as the overall result. Final results are obtained by majority voting and are used for comparison with Gemini-based evaluation.

\paragraph{Gemini Prompt.}
The Gemini prompt used for CoT plausibility evaluation is shown below.

{\footnotesize
\begin{quote}
\textbf{System role:} You are an expert in Chinese dialogue speech style analysis.

\textbf{Task:} Determine whether the ``model-predicted CoT'' is reasonable based on the target speaker profile, dialogue history, the current target utterance text, and multiple audio segments provided in order.

\textbf{Goal:} Your goal is not to check whether the prediction exactly matches a unique ground-truth answer, but whether it can be considered a reasonable, natural, and context-consistent interpretation of the current target utterance and its audio.

\textbf{Audio order:}
The first several audio segments correspond to historical dialogue turns. The last audio segment corresponds to the current target utterance. Your judgment should focus on whether the predicted CoT for the current target utterance is reasonable, while also considering dialogue relations, emotional flow, and tone continuity from the history.

\textbf{CoT dimensions:}
emotion: whether the predicted emotion of the current target utterance is reasonable;
tone: whether the predicted tone / discourse function is reasonable;
pitch: whether the predicted pitch trend is reasonable;
energy: whether the predicted energy / loudness is reasonable;
speed: whether the predicted speaking rate is reasonable;
overall: true only if all five dimensions are reasonable; otherwise false.

\textbf{Judging principles:}
If a dimension clearly conflicts with the perceptual impression of the current target utterance audio, output false. If a dimension clearly conflicts with the dialogue relation implied by the history, output false. If a dimension is broadly consistent with both the audio and the history, output true. It may differ from human annotation and still be true, as long as it is a reasonable interpretation. Ignore factors unrelated to CoT, such as audio quality, naturalness, minor mispronunciations, or ASR deviations. Prioritize the current target utterance audio, and then use the dialogue history as supporting context.

\textbf{Output format:}
Please output strict JSON only, without any extra explanation:
\{
``emotion'': true,
``tone'': false,
``pitch'': true,
``energy'': false,
``speed'': true
\}

\textbf{Sample to be evaluated:}
\texttt{<Insert sample here>}
\end{quote}
}

\subsubsection{Additional Human Evaluation Details}

To further improve the reliability of our experimental results, we provide additional details of the human evaluation setup and score aggregation procedure.

For single-utterance human evaluation, we randomly sampled 45 test cases and invited 25 annotators to independently rate the outputs of each model. The evaluation dimensions include overall description consistency, identity consistency, timbre consistency, expressive-style consistency, role-intent consistency, control stability, and MOS naturalness. All dimensions were rated on a 1--5 scale, and the final score of each model was obtained by averaging the ratings across annotators and samples.

For CoT prediction evaluation in dialogue, in addition to Gemini-based automatic evaluation, we further conducted human verification on the 30-sample subset described in Section~B.2.1, with 15 annotators independently judging whether the predicted CoT was reasonable. The annotators provided binary judgments for each of the five CoT attributes as well as the overall result, and the final decision was obtained by majority voting.

For overall dialogue evaluation, we randomly sampled 15 dialogue cases and invited 20 annotators to perform human pairwise comparison under the same dialogue context. For each case, we synthesized the full dialogue session for both the full model and the w/o-CoT variant. The target speaker was selected from \texttt{eligible\_target\_speakers}, while the other speaker was kept as dialogue history/reference. Annotators then judged which system produced target-speaker turns that were more coherent with the full session context. For pairwise comparisons such as w/ CoT versus w/o CoT, the presentation order was randomized to reduce position bias, and the final preference was determined by majority voting. We report the resulting human pairwise preference on context coherence.

\section{Analysis of Hierarchical Variational Timbre Conditioning}                                                                                                                          
  \label{sec:appendix-timbre-reuse}                                                               
  This appendix provides additional empirical analysis for the factorized hierarchical variational timbre conditioning module proposed in Section~3.3. The central claim of that module is    
  that the utterance-level speaker embedding $e_{\text{spk}}$, together with the KL regularization between the segment posterior $q(z_2 \mid s)$ and the utterance-conditioned prior $p(z_2
  \mid e_{\text{spk}})$, enables a desired timbre to be \emph{preserved and reused} across diverse utterances while keeping expression adaptive to the local context. Here we directly probe  
  this property and use it to interpret the human-evaluated timbre score of CapTalk in Table~2.

  The human-evaluated timbre consistency reported in Table~2 (3.59 for CapTalk) is lower than that of several baselines. We attribute this not to a deficiency of the hierarchical variational
   module, but to a structural property of caption-conditioned voice design that is often overlooked in evaluation: voice design is intrinsically a one-to-many task. Given a caption such as
  ``a calm middle-aged man with a low and steady voice'', many concrete voices are equally valid realizations. In our human evaluation, each generation independently samples a new           
  $\hat{e}_{\text{spk}}$ from the caption-conditioned distribution, so different samples for the same caption may correspond to different specific voices within the same caption-consistent
  voice space. Annotators tend to perceive such sample-level variability as timbre inconsistency, which deflates the timbre score even when the module itself preserves timbre faithfully
  under a fixed designed voice.

  To probe whether the hierarchical variational timbre conditioning module actually delivers the timbre-reuse property claimed in Section~3.3, we conduct a \emph{timbre reuse} analysis that 
  directly contrasts the two inference modes supported by the module. For each caption, we evaluate two settings: (i) \emph{Fixed $e_{\text{spk}}$}, where $\hat{e}_{\text{spk}}$ is predicted
   once from the caption and then reused across all subsequent utterances, corresponding to the practical voice-design-then-reuse usage enabled by the module; and (ii) \emph{Resampled       
  $e_{\text{spk}}$}, where $\hat{e}_{\text{spk}}$ is independently re-predicted for every utterance, corresponding to the protocol implicitly assumed by per-sample human rating. For each
  setting, we generate multiple utterances with diverse text content, extract speaker embeddings from the generated speech using a pretrained WavLM model~\cite{chen2021wavlm}, and compute
  the average pairwise cosine similarity (SIM) across utterances of the same designed voice. Higher SIM indicates stronger cross-utterance timbre consistency. We emphasize that the
  per-sample human evaluation in Table~2 corresponds to the \emph{Resampled} setting, not the \emph{Fixed} setting: each rated sample uses an independently re-predicted
  $\hat{e}_{\text{spk}}$. The 3.59 timbre score should therefore be read against the 0.42 SIM in the resampled mode, while the 0.92 SIM in the fixed mode reflects a complementary inference
  path of the same module that human raters never accessed.

  \begin{table}[h]
    \centering
    \small
    \caption{Timbre reuse analysis. SIM is the average pairwise cosine similarity between WavLM speaker embeddings extracted from generated utterances of the same designed voice.}
    \label{tab:appendix-timbre-reuse}                                                                                                                                                         
    \begin{tabular}{lc}                                                                                                                                                                       
      \toprule                                                                                                                                                                                
      Setting & Cross-utterance SIM $\uparrow$ \\                                                                                                                                             
      \midrule                                              
      Fixed $e_{\text{spk}}$ (timbre reuse) & 0.92 \\                                                                                                                                         
      Resampled $e_{\text{spk}}$ (per-utterance) & 0.42 \\
      \bottomrule                                                                                                                                                                             
    \end{tabular}                                           
  \end{table}                                                                                                                                                          
                                                            
  As shown in Table~\ref{tab:appendix-timbre-reuse}, the fixed-$e_{\text{spk}}$ setting achieves a cross-utterance SIM of 0.92, substantially higher than the 0.42 obtained under the         
  resampled setting, confirming that once a designed voice is fixed, the hierarchical variational timbre conditioning module preserves it consistently across utterances. The lower timbre
  score in Table~2 therefore primarily reflects the inherent variability of caption-conditioned sampling rather than a weakness in timbre modeling. This observation also suggests that future
   evaluation protocols for voice design should explicitly distinguish two complementary axes: \emph{caption-conditioned diversity}, which characterizes how broadly the model covers the
  space of caption-consistent voices, and \emph{designed-voice reuse consistency}, which characterizes how stably a single designed voice can be preserved across utterances and dialogue
  contexts. The hierarchical variational timbre conditioning module proposed in Section~3.3 is specifically aimed at the latter, and the timbre reuse analysis above provides direct empirical
   evidence of its effectiveness, complementing the architectural motivation given in the main text.

\section{Examples of Single-Utterance and Dialogue Captions}
\label{sec:appendix-caption-examples}

Although we generate the same three surface styles (APS, DSD, and RP) in both the single-utterance and dialogue settings, the semantic scope of the caption is fundamentally different. In the single-utterance setting, the caption is tied to one specific recording and is intended to describe how that utterance is spoken, including its local expressive realization. In the dialogue setting, the caption is used as a reusable speaker condition attached to the target speaker across many turns in the same session. The transient turn-level state of the current utterance is instead represented separately by the CoT sequence described in Section~4.2. In this sense, the single-utterance caption mainly answers ``how is this utterance spoken?'', whereas the dialogue caption mainly answers ``what kind of speaker is this across the dialogue?''.

Table~\ref{tab:appendix-caption-comparison} provides representative examples of the three caption styles used in our data construction. For APS captions, we further note an important difference between the two settings. In the single-utterance setting, APS captions may include utterance-level expressive information. In the dialogue setting, however, the speaker-level APS caption is designed to capture relatively stable speaker traits only. Therefore, emotion and tone are not extracted into the dialogue APS caption; instead, they are modeled separately by the CoT sequence together with other turn-level dynamic attributes. This separation helps avoid mixing persistent speaker identity with transient turn-wise expression in the same caption.

\begin{table*}[t]
  \centering
  \footnotesize
  \caption{Illustrative comparison between single-utterance and dialogue captions.}
  \label{tab:appendix-caption-comparison}
  \begin{tabular}{>{\raggedright\arraybackslash}p{1.4cm}>{\raggedright\arraybackslash}p{6.3cm}>{\raggedright\arraybackslash}p{6.3cm}}
    \toprule
    Style & Single-utterance caption & Dialogue caption \\
    \midrule
    APS &
    Gender: Female. Age: Teenager. Pitch: Female pitch range, relatively high at the beginning, naturally falling in the middle, and slightly rising at the end of the sentence. Speaking rate: The overall speaking rate tends to be relatively fast, with a compact rhythm and frequent local speed changes. Volume: The overall loudness is moderate, but some segments within the utterance are relatively louder, with a moderate dynamic range. Clarity: Clear articulation, standard pronunciation, and complete final consonants. Fluency: Fluent delivery, with no obvious pauses or filler words. Accent: Standard Mandarin. Timbre texture: Bright timbre with a slight childlike quality, soft and warm. Personality: Lively, cheerful, good at guiding others, and approachable. &
    Gender: Male. Age: Middle-aged. Pitch: The pitch center tends to be moderately low, with strong F0 stability and solid chest resonance. Speaking rate: The overall speaking rate tends to be medium-slow, with a steady rhythm, though it becomes slightly faster in some narrative segments. Volume: The overall loudness is moderate, with a relatively narrow dynamic range and little loudness variation across segments. Clarity: Clear articulation, accurate placement, and complete final consonants. Fluency: Fluent and coherent speech, with occasional minor discourse fillers. Accent: A noticeable northern accent, with relatively hard pronunciation and flatter intonation. Timbre texture: Slightly rough voice quality with mild graininess, with resonance concentrated in the oral and chest cavities. Personality: Easygoing and natural, with a willingness to narrate and a tendency to share reflections from everyday life. \\
    DSD &
    This is a young adult male voice with a somewhat dark and slightly husky timbre, as if he has just woken up or is in a low mood. He speaks at an unhurried pace, giving the impression of being immersed in his own memories. &
    This voice sounds like that of a middle-aged northern man with some life experience. His timbre is steady and slightly husky, and he speaks at an unhurried pace with an easygoing manner. He sounds like someone who enjoys chatting with others and remains curious about new things. \\
    RP &
    An experienced team leader is reminding the group to prepare mentally for the high-intensity challenge ahead. The tone is solemn and authoritative, emphasizing the importance of focus and psychological readiness. &
    An experienced elder woman is patiently analyzing a complicated interpersonal relationship. Her speech carries both life experience and practical reminders, and her tone is gentle yet firm, as if she were guiding a younger person through confusion. \\
    \bottomrule
  \end{tabular}
\end{table*}

Two observations are worth highlighting. First, even when the surface style is matched (APS-to-APS, DSD-to-DSD, and RP-to-RP), single-utterance captions still contain more local information about the current recording, while dialogue captions are comparatively better suited to reusable speaker-level characterization. Second, dialogue generation requires a cleaner separation between stable speaker traits and turn-wise expressive variation. This is why we pair speaker-level captions with CoT rather than asking a single caption to simultaneously encode both long-term speaker identity and the short-term state of every turn.

\end{document}